\documentclass[aps, prb, reprint, amsmath, amssymb, superscriptaddress]{revtex4-1}

\usepackage{subfigure}
\usepackage{natmove}
\usepackage{dcolumn}
\usepackage{graphicx}
\usepackage{wrapfig}
\usepackage{color}
\usepackage{bm}
\graphicspath{{pictures/}}
\usepackage{notoccite}
\usepackage{lipsum}
\usepackage{mathptmx}
\begin{document}

\title{Computational Synthesis of Substrates by Crystal Cleavage}

\author{Joshua T. Paul}
\affiliation{Department of Materials Science and Engineering, University of Florida, Gainesville, Florida 32611, USA}
\affiliation{Quantum Theory Project, University of Florida, Gainesville, Florida 32611, USA}
\author{Alice Galdi}
\affiliation{Department of Physics, Cornell University, Ithaca, New York 14850, USA}
\affiliation{Cornell Laboratory for Accelerator-Based Sciences and Education, Cornell University, Ithaca, New York 14850, USA}
\author{Christopher Parzyck}
\affiliation{Department of Physics, Cornell University, Ithaca, New York 14850, USA}
\author{Kyle Shen}
\affiliation{Department of Physics, Cornell University, Ithaca, New York 14850, USA}
\author{Richard G. Hennig}
\email{rhennig@ufl.edu}
\affiliation{Department of Materials Science and Engineering, University of Florida, Gainesville, Florida 32611, USA}
\affiliation{Quantum Theory Project, University of Florida, Gainesville, Florida 32611, USA}

\date{\today}

\begin{abstract}
The discovery of novel substrate materials has been dominated by trial and error, opening the opportunity for a systematic search. To identify stable crystal surfaces, we generate bonding networks for materials from the Materials Project database with one to five atoms in the primitive unit cell. For three-dimensional crystals in this set, we systematically break up to three bonds in the bonding network of the primitive cell. Successful cleavage reduces the bonding network to two periodic dimensions, creating a layer of the cleaved crystal. We identify 4,693 unique cleavage surfaces across 2,133 bulk crystals, 4,626 of which have a maximum Miller index of 1. To characterize the likelihood of cleavage and the thermodynamic stability of the cleaved surfaces, we create monolayers of these surfaces and calculate the work of cleavage and the partially-relaxed surface energy using density functional theory to discover 3,991 potential substrates, 2,307 of which do not contain {\it f}-valence electrons and 2,183 of which are derived from a bulk precursor with an entry in the Inorganic Crystal Structure Database. Following, we identify distinct trends in the work of cleavage of these layers and relate them to metallic and covalent/ionic bonding of the three-dimensional precursor. We also assembled a database of commercially available substrates and show that the database of predicted substrates significantly enhances the diversity and range of the distribution of electronic properties and lattice parameters, providing opportunities for the epitaxial growth of many materials. We illustrate the potential impact of the substrate database by identifying several new epitaxial substrates for the transparent conductor BaSnO$_3$, which exhibit low cleavage energies and result in strains an order of magnitude lower than currently used substrates.
The open-source database of predicted and commercial substrates and their properties is available at MaterialsWeb.org.

\end{abstract}

\maketitle

\section{Introduction}

Single-crystal substrates facilitate the epitaxial growth of crystalline thin films based on their surface similarity, in terms of lattice parameters and symmetry. Many single-crystal substrate materials are commercially available with different facet orientations. However, while exfoliated 2D materials such as graphene are naturally smooth, cleaving a crystal does not guarantee an atomically smooth surface. Energetic instabilities of cleaved facets can result in the formation of various defects that limit the quality of a substrate for synthesis efforts. Identifying a wider range of substrates with low surface energies would enable the growth of high-quality materials by minimizing the presence of undesirable defects. For example,  the cubic perovskite BaSnO$_3$ has diminished electronic properties due to the lattice mismatch of its (100) surface and the substrates suitable for its growth~\cite{paik2017, lee2017_per, Yoon2018}. In addition to lattice mismatch, one must consider the reactivity between a substrate and the thin film being grown, for both the precursor chemical reactions and the final thin film composition, which will further reduce the number of available options.

\begin{figure}[tb]
    \centering
    \includegraphics[width=\columnwidth]{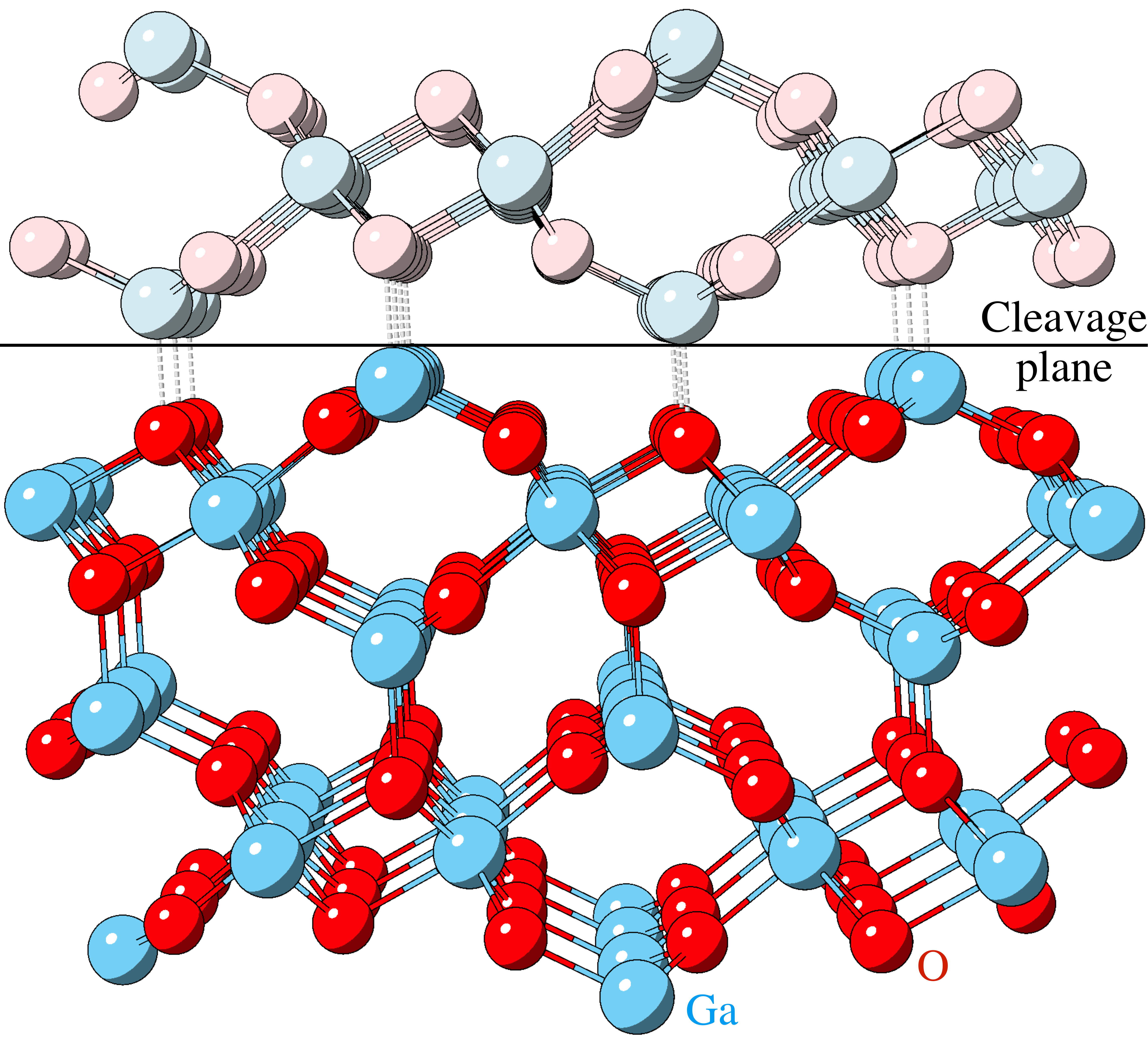}
    \caption{Example of a cleavable crystal, Ga$_2$O$_3$. The (20$\overline{1}$) plane exhibits a low density of bonds that, when cleaved, creates a low energy surface.}
    \label{fig:1}
\end{figure}

The discovery of two-dimensional (2D) materials present similar challenges to substrates, most notably that both require materials with low surface energies. Computational efforts to identify novel 2D material have significantly expanded the list of potential monolayers and helped guide experimental synthesis~\cite{sahin2009, zhuang2013, Lebeque2013, ashton2017, Choudhary2017, cheon2017, Mounet2018, Haastrup2018}.
One discovery technique is data mining, which searches bulk materials databases for yet unidentified monolayers~\cite{Lebeque2013, ashton2017, cheon2017, Choudhary2017, Mounet2018, Haastrup2018}. A recent effort in this direction identified and characterized the bonding network of a crystal structure using the topological scaling algorithm (TSA)~\cite{ashton2017}. This algorithm identifies bonding clusters within a finite number of unit cells, then uses the scaling of this network size as a function of supercell size to define dimensionality. One of the most significant contributions of this approach is that one does not require one to define a facet to search along {\it a priori}. The algorithm itself will identify a monolayer with a surface parallel to any facet cut, and thus renders the issue of orientation mute. This resulted in the discovery of over 600 low-energy 2D materials~\cite{ashton2017} and the creation of the MaterialsWeb.org 2D material database.


Previous computational efforts have been made to identify potential substrate materials, e.g., Ding et al.~\cite{Ding2016} searched the Materials Project database for a substrate to stabilize a metastable polymorph of VO$_2$. Using lattice mismatch and induced thin-film strain as criteria, they identify several candidates. That search considered a set of low-index facets but did not account for the surface energy of the facets. This illustrates the need for an algorithm to identify likely substrates and a database of low-energy work of cleavage facets, which is the scope of this work.


In this work, we present an approach to identify novel substrates, which utilizes the computational cleavage of bulk crystals. This effort is motivated by our recent 2D structure search using the genetic algorithm software GASP~\cite{Tipton_2013, Revard2016} for the Ga$_{2}$O$_{3}$ system. In that search, we identified a monolayer structure that can be cleaved from the bulk crystal and was already experimentally synthesized~\cite{Hwang2014Ga2O3, Ga2O3_og}. Thus, we developed data mining techniques to discover planes of cleavage in fully periodic crystals. Rather than using the TSA to identify van der Waals gaps, we use it first to identify conventionally bonded solids; then, we systematically break bonds in the crystal to create a low-dimensional structure. When the breaking of bonds generates a surface (i.e., gives the material a 2D structural motif), we extract a unit cell thick monolayer and calculate the work of cleavage (i.e., the energy cost required to cleave the crystal and form two surfaces) using density functional theory (DFT). This approach identifies nearly 4,000 potential substrates with surface energies comparable to experimentally used substrates. 

\section{Results}
{\it Candidate materials.} For our study, we select all materials in the  MaterialsProject database~\cite{jain2013} with five or fewer atoms in the primitive unit cell and within 50~meV/atom of the convex hull, the latter to promote thermodynamic stability of the final surface. Next, the topological scaling algorithm (TSA) identifies conventionally networked structures from this subset and creates a list of all bonds between neighboring atoms in these crystals. The bonding identification utilizes the empirical atomic radii of the elements, as provided in the pymatgen software package~\cite{ong2013}, which we increase by 10\%.

\begin{figure}[tb]
    \centering
    \includegraphics[width=\columnwidth]{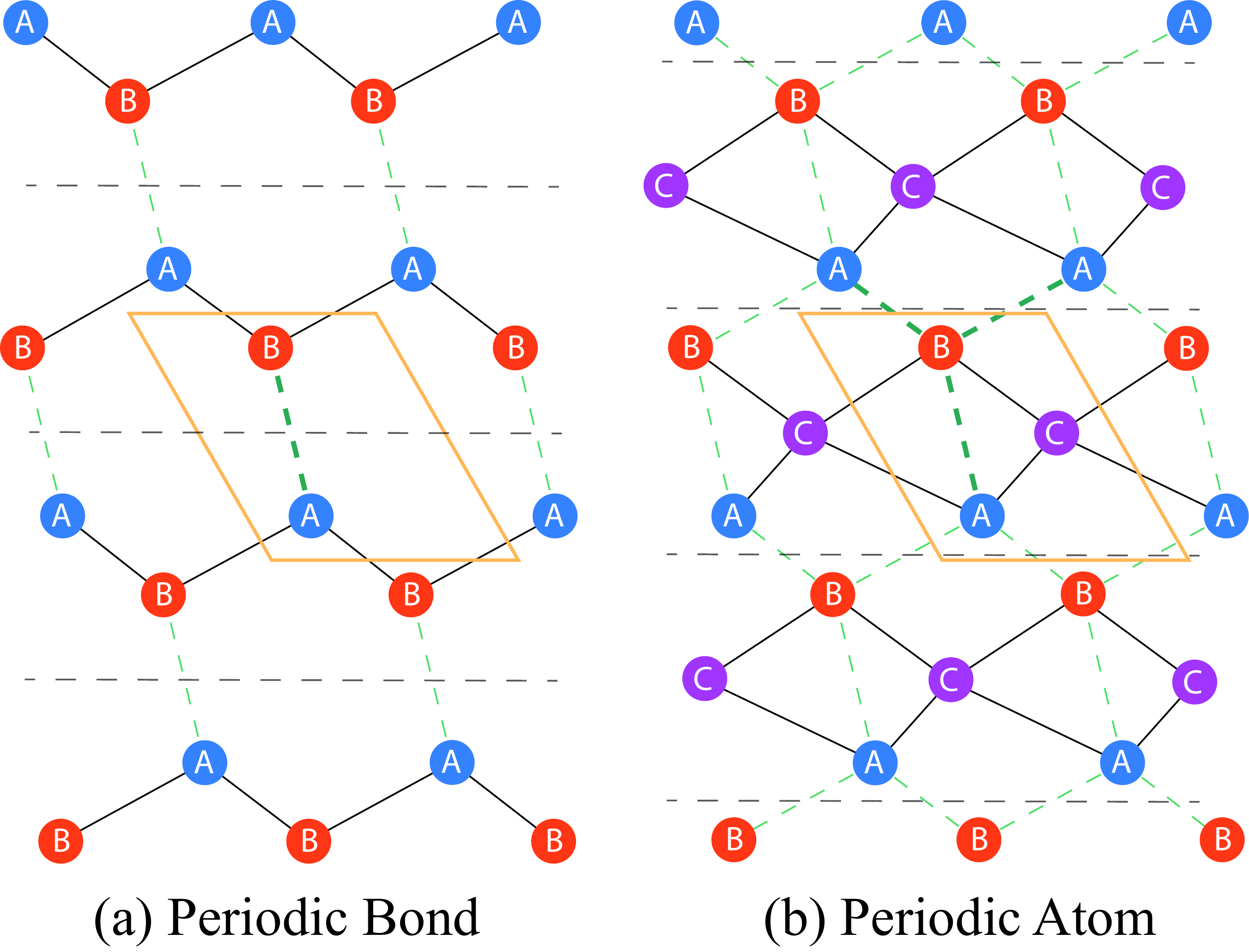}
    \caption{Bond cleaving algorithm. (a) The periodic bond approach cleaves all periodic images of a bond. (b) The periodic atom approach cleaves the bonds between all periodic images of the atoms. The periodic atom approach requires at least three atoms in the primitive cell to generate a cleaved facet.}
    \label{fig:2}
\end{figure}

{\it Bond cleavage.} We implement two approaches for cleaving bonds in a crystal structure, which we illustrate in Fig.~\ref{fig:2}. The first approach breaks all periodic instances of a bond between two atoms in the supercell and is denoted as the ``periodic bonds" approach. In this case, the three A-B bonds shown in Fig.~\ref{fig:2}(a) are treated separately. The second approach breaks the bonds between all periodic instances of the atoms in the supercell and is referred to as the ``periodic atom" approach, illustrated in Fig.~\ref{fig:2}(b).

We systematically break unique bonds in the primitive cell up to a maximum number of bonds given by
\begin{equation}
	\label{eq:1}
	N_\mathrm{bonds} = \mathrm{round} \left ( \alpha\,N_\mathrm{atoms}^{2/3} \right ),
\end{equation}
where $N_\mathrm{atoms}$ is the number of atoms in the primitive cell, and $\alpha$ is a scalable parameter based on the desired maximum number of bonds broken. The exponent of 2/3 appropriately scales the number of bonds per plane with increasing cell size. The choice of $\alpha=1$ in this work results in one to three cleaved bonds in primitive cells of one to five atoms. For the periodic bond approach, $N_\mathrm{bonds}$ scales with the square of the supercell size. For the periodic atom approach, $N_\mathrm{bonds}$ scales at least with the square of the supercell size, with the potential to break a significantly larger number of bonds than the periodic bond approach. 

These two approaches are complementary, as illustrated in Fig.~\ref{fig:2}. For example, applying the periodic atom approach to the structure in Fig.~\ref{fig:2}(a) or the periodic bond approach to the structure in Fig.~\ref{fig:2}(b) would both fail to cleave a surface with $\alpha=1$ in Eq.~\ref{eq:1}. As the unit cell size increases, the periodic bond and periodic atom approaches become equivalent. With a high enough $\alpha$ value, the periodic bond approach will also identify the cleaved surfaces of the periodic atom approach. However, this would be significantly more computationally expensive due to the increase in possible combinations of broken bonds, which need to be considered.


\begin{figure}[b]
    \centering
    \includegraphics[width=\columnwidth]{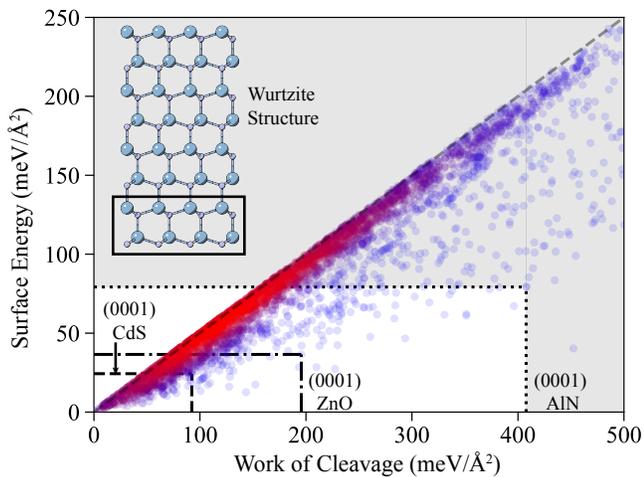}
    \caption{The work of cleavage and the surface energy after ten relaxation steps of the cleaved surfaces. Red indicates a higher density of points. The solid, dashed, and dotted black lines represent the work of cleavage of the (0001) CdS, ZnO, and AlN substrates, respectively. The grey region indicates materials have a work of cleavage greater than common substrates.}
    \label{fig:3}
\end{figure}


\begin{figure*}[t]
    \includegraphics[width=\textwidth]{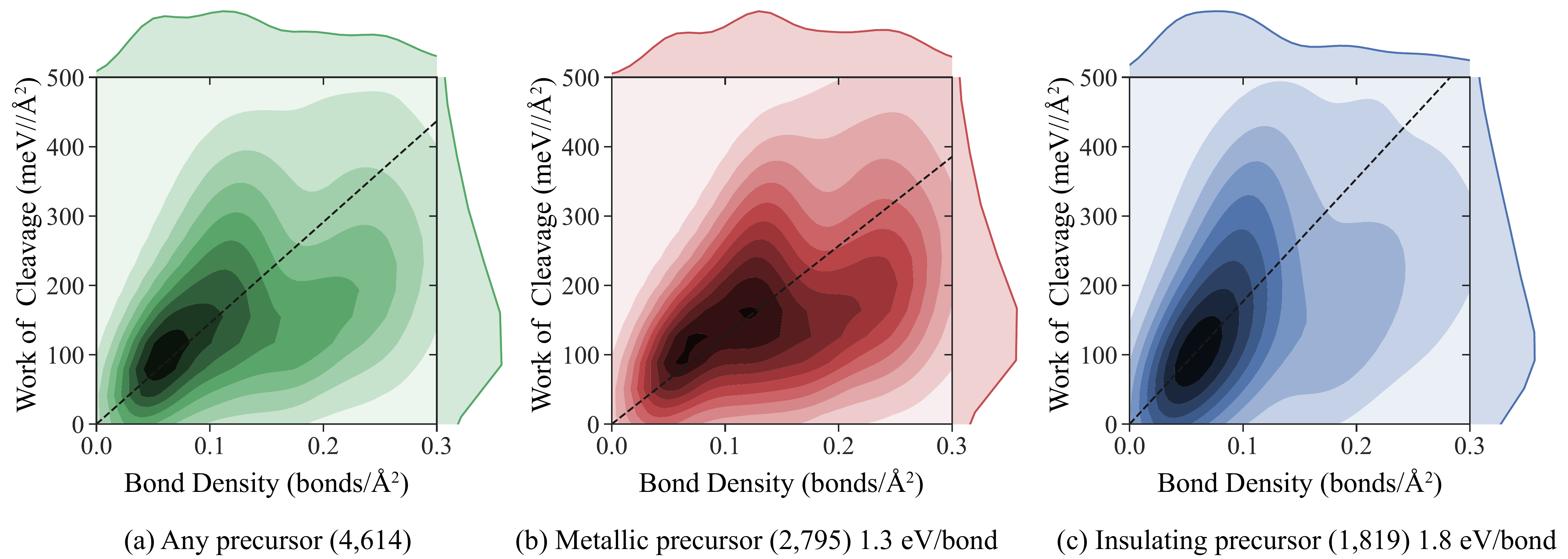}
    \caption{Distribution of the work of cleavage vs.\ bond density of cleaved surfaces, represented by kernel density estimation: (a) for 4,614 cleavage surfaces, (b) for surfaces with metallic precursors, and (c) for surfaces with precursors exhibiting an electronic bandgap. We use the bandgap reported by Materials Project for the bulk precursors of the surfaces. The dashed lines indicate the mean energy per bond. On average, the metallic precursors exhibit a higher coordination number, and hence bond density for the cleavage plane in conjunction with a lower energy per bond than the more covalent and ionically bonded precursors. This trend is reflected in the average energy of the cleaved bonds for metals being lower at 1.3~eV/bond compared to 1.8~eV/bond for the cleaved covalent and ionic precursors.
    }
    \label{fig:4}
\end{figure*}

{\it Topology of resulting structure.} To determine if a cleavage surface has been generated, the TSA is run on the three-dimensional primitive cell with $N_\mathrm{bonds}$ or fewer bonds broken (using either approach). If the TSA identifies a two-dimensional structural motif with the same stoichiometry as the overall cell, we have created a cleavable surface. This surface is isolated as a unit cell thick monolayer and oriented such that the $\vec a$ and $\vec b$ lattice vectors span the 2D lattice of the monolayer structure and the $\vec c$ lattice parameter is chosen perpendicular to the $(\vec a, \vec b)$ plane. Due to the crystal symmetry, our algorithm can identify multiple instances of some surfaces. We remove these duplicates and identify the unique surfaces extracted from each crystal using the pymatgen structure matching algorithm~\cite{jain2013}.

{\it Predicted substrates.}
From a starting set of 120,612 materials in the Materials Project database~\cite{jain2013}, 9,089 crystals meet the criteria of exhibiting (i) a formation energy within 50 meV/atom of the thermodynamic hull (72,316), (ii) a primitive cell of 5 or fewer atoms (9,980), (iii) all lattice angles greater than 16$^{\circ}$ and less than 164$^{\circ}$ (9,973), and (iv) a bonding network of three-dimensional topology. Criteria (iii) is introduced as a result of the exfoliation algorithm, which is found to return incorrect cleaved surfaces if the lattice has an extreme lattice angle. Though applying the algorithm to the conventional cell of these crystals resolves the issue, this results in a unit cell with greater than 5 atoms and thus are not considered. Applying the periodic bond and periodic atom cleavage approaches to the 9,089 materials generates 1,925 and 4,006 unique surfaces, respectively. We apply the pymatgen structure tool to the combined list of 5,931 surfaces to identify a total of 4,693 unique cleaved surfaces across 2,133 bulk crystals.
respectively. We apply the pymatgen structure tool to the combined list of 5,845 surfaces to identify a total of 4,693 unique cleaved surfaces across 2,133 bulk crystals. It is worth noting here that Ga$_2$O$_3$ is not identified as cleavable by this search, as (i) the number of atoms in the primitive cell of the crystal is greater than five and (ii) the TSA identifies the crystal as layered when using our choice of 1.1 times the atomic radii of Ga and O, rather than as a fully networked structure.

To determine the stability of the surfaces, we perform DFT calculations with VASP~\cite{kresse1996vasp,bloche1994paw}. We extract a single unit cell thick slab for each of the identified 4,693 unique cleavage surfaces and perform geometric optimization of atomic positions within the slabs and a single-point calculation for the bulk precursor to maintain consistency across our energy comparisons. Of the 4,693 monolayers, 4,614 converged or reached our maximum number of iterations with a final surface energy of at most half the work of cleavage. A sizable number of 2,015 of these surfaces contain {\it f}-valence elements, for which semilocal exchange-correlation functionals exhibit larger formation energy errors~\cite{kirklin2015}. Therefore, care should be taken when considering these materials in the MaterialsWeb.org substrate database.

We use the work of cleavage of a monolayer, $E_\mathrm{cleave}$, to measures the energy required to cleave a unit area of the bulk precursor material and the surface energy, $E_\mathrm{surf}$, to describe the thermodynamic stability of a material’s facet:
\begin{equation}
    \label{eq:ad}
    E_\mathrm{cleave} = \frac{(E_\mathrm{sub,as-cut} - E_\mathrm{bulk} \, \frac{N_\mathrm{sub}}{N_\mathrm{bulk}})}{A_\mathrm{sub}}
\end{equation}
and 
\begin{equation}
    \label{eq:surf}
    E_\mathrm{surf} = \frac{(E_\mathrm{sub,opt} - E_\mathrm{bulk} \, \frac{N_\mathrm{sub}}{N_\mathrm{bulk}})}{2 A_\mathrm{sub}},
\end{equation}
where $E_\mathrm{sub,as-cut}$ and $E_\mathrm{sub,opt}$ are the energies of the as-cut substrate immediately after cleaving and of the optimized substrate after relaxing the atomic positions; $E_\mathrm{bulk}$ is the energy of the bulk precursor, $N_\mathrm{bulk}$ and $N_\mathrm{sub}$ represent the number of atoms in the bulk and substrate, respectively, and $A_\mathrm{sub}$ denotes the substrate area. The factor of two difference is due to the work of cleavage being a measure of the energy needed to create two surfaces, while the surface energy is the thermodynamic stability of a single surface.

To validate that a monolayer accurately approximates the work of cleavage for cleaving a crystal, we calculate the change in work of cleavage with slab thickness for a subset of 21 (001) surfaces, from one to four unit cells. We find changes in the work of cleavage of less than 3~meV/\AA$^2$ in these systems, confirming that one unit cell thick slabs sufficiently describe the work of cleavage for a cleaved surface in this high-throughput effort. To verify that the work of cleavage indicates surface stability, we compare the calculated work of cleavage for 4,614 cleaved, free-standing slabs to their partially-optimized surface energy in Fig.~\ref{fig:3}. We observe that 2,557 materials display surface energies within 10\% of half their work of cleavage and 1,615 within 5\%. 

We note that there is the possibility of surface reconstructions for some of the predicted substrates. However, due to the high computational cost, we do not perform a search for  possible reconstructions. The thermodynamic driving force for surface reconstructions is the surface energy. Hence, the probability for reconstructions decreases for lower energy surfaces. 
However, before pursuing synthesis of a substrate in this work, we recommend determining at least the dynamic stability of a surface structure using phonon calculations and, if possible, searching for reconstructions using, e.g., genetic algorithm searches~\cite{Revard2016}. Note that as this is a global optimization problem~\cite{revard2014Book}, there is no guarantee that any reconstruction, let alone the most stable reconstruction, will be identified with computational techniques. Reconstructions will also be significantly altered by the adsorption of species in the first stages of the film growth on the substrate, making this problem more closely related to the specific film-substrate interface.

\section{Discussion}
Our search identifies three surfaces currently used as substrates: hexagonal CdS, ZnO, and AlN. These crystals share the same structure and are all commercially available as (0001) substrates. In our search, we identify four substrates for each material: one (1$\overline{1}$0), one ($\overline{1}$01), and two (001) facets. These facets are equivalent to the Miller-Bravais convention of (1$\overline{1}$00), ($\overline{1}$011), and (0001). The most stable facet for these crystals is (1$\overline{1}$0), with a work of cleavage of 56, 123, and 275 meV/\AA$^{2}$ for CdS, ZnO, and AlN, respectively. The most stable (001) facet of each crystal has a work of cleavage of 92, 196, and 407~meV/\AA$^{2}$, respectively. Fig.~\ref{fig:3} highlights 3,991 surfaces (2,307 {\it f}-valence free and 2,183 with an entry in the Inorganic Crystal Structure Database (ICSD) entry) with a work of cleavage below that of (001) AlN, and 935 (718 {\it f}-valence free and 418 with an ICSD entry) surfaces below that of (001) CdS. 

We consider any slab with a work of cleavage less than (001) AlN to be a suitable substrate and included in our substrate database. Some facet cuts could have different terminations for the same cleaved unit. For example, the cleaved (001) AlN can has both an aluminum-terminated and nitrogen-terminated facet. For these cases, the reported work of cleavage and surface energy are an average of these terminations. In addition to the energy benchmarks, we identify the conventional cell hkl index of the facets we cut. Of the 4,693 facets, 4,626 have Miller indices no greater than 1, showing that the vast majority of planes we cleave are low-index facets. Of the remaining facets, 63 have a maximum Miller index of 2 and 4 have a maximum Miller index of 3. While our choice of primitive cells biases the indices found, these results demonstrate how our approach is agnostic to the orientation of the facet cuts and the potential for future searches to expand this substrate database.

\begin{figure*}[t]
    \includegraphics[width=\textwidth]{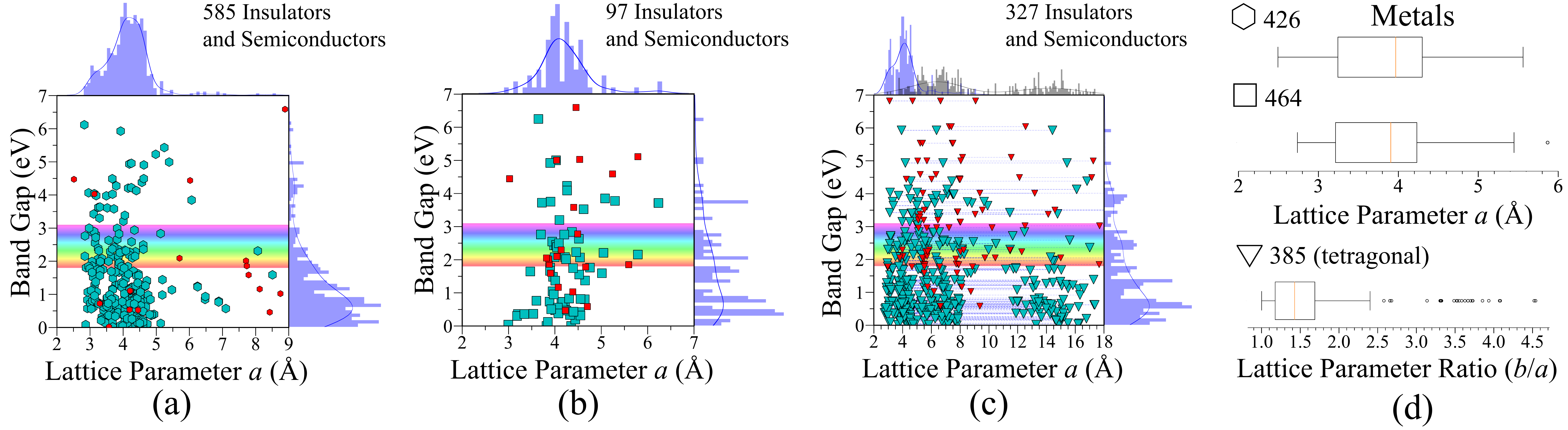}
    \caption{Lattice vectors and bandgaps for (a) hexagonal, (b) square, and (c) tetragonal surfaces of semiconducting or insulating substrates identified in this work. For the tetragonal substrates in (c), the {\it a} and {\it b} lattice vectors are both plotted and connected with a dashed line, and the {\it b} kernel density is plotted in grey. 93 of the 106 commercially available insulating substrates identified are shown in red filled markers. For metallic substrates, (d) shows boxplots for the hexagonal and square surface structures and the {\it b}/{\it a} ratio of the tetragonal ones. The 23 monoclinic surfaces are not represented in these figures. Surfaces are classified with a maximum lattice vector difference of 1\% and a maximum in-plane angle difference of 0.5$^{\circ}$. Substrates containing $f$-valence elements are not shown.
    }
    \label{fig:5}
\end{figure*}

Fig.~\ref{fig:4} shows the trend between the number of cleaved bonds and the work of cleavage across the 4,614 converged cleaved surfaces. The joint distribution in Fig.~\ref{fig:4}(a) indicates that our cleavage criterion of Eq.~\eqref{eq:1} results in both a low density of cleaved bonds and a moderate spread of work of cleavage. Furthermore, the distribution indicates two clustering trends in the plot, which correspond to different average bond energies. We attribute these trends to different types of chemical bonds being broken. Metallic systems typically display high coordination numbers and somewhat lower bond energies. Covalent and ionic bonds share localized electrons, which results in lower coordination numbers and higher energies per bond. 

To test this hypothesis, Figures~\ref{fig:4}(b) and (c) show the work of cleavage and bond density distribution for monolayers derived from metallic and insulating precursors, respectively, based on the precursor bandgap reported in the MaterialsProject database~\cite{jain2013}. We observe that the two clusters in the distribution indeed correspond predominantly to metallic and covalent/ionic bonding. Metallic bonds typically occur in materials with higher coordination numbers and have lower average energy than covalent or ionic bonds, resulting in a broad spread of bond densities in Fig.~\ref{fig:4}(b). In contrast, covalent/ionic bonds typically correspond to lower coordination numbers with higher energies for each bond, which is reflected in the distribution in Fig.~\ref{fig:4}(c). The average bond energies of the metallic and insulating precursors are 1.3 and 1.8~eV/bond, respectively, further demonstrating the difference of bonding in these systems. As bonds exist across a continuous spectrum of metallic, covalent, and ionic character, the average bond energy is not a precise description of the bonding in these systems. However, the observed bond energies in the metallic and insulating precursor materials reveal an overall consistent trend.

To determine the potential of our data mining approach to expand the set of known substrates, we characterize the cleavage surfaces' symmetry and lattice parameters. Fig.~\ref{fig:5} illustrates the lattice parameter distribution of the hexagonal, square, and tetragonal substrates identified in this work which are free of $f$-valence elements, as well as the electronic properties of their bulk precursors. The broad range of electronic behavior and lattice parameters for each symmetry indicates that these cleaved crystals could provide suitable substrates for a variety of thin film systems. 

We follow by creating a database of commercially available substrates totaling 190 substrates and include the data in Fig.~\ref{fig:5} using crystal structures and band gaps sourced from the Materials Project database. Comparing the database of predicted substrates to the one of commercially available substrates shows two significant advancements.  First, we greatly expand the list of substrates exhibiting band gaps within the visible spectrum and beyond, creating more opportunities for optical excitation of thin films from beneath the substrate rather than from above. Second, we identify substrates with a broader range of lattice parameters, which can help synthesis efforts. For example, currently, 23 commercially available hexagonal substrates exhibit lattice parameters between 2.5 and 5.0~\AA. The computational database expands this list to 984 hexagonal substrates within the same range. This increase provides more opportunity to identify a substrate that both epitaxially matches the growth crystal and is chemically compatible.

To demonstrate the application of this substrate database, we epitaxially match cubic perovskite (100) BaSnO$_3$, using the crystal structure and stiffness tensor provided by MaterialsProject~\cite{jain2013, DeJong2015} to  substrates with a work of cleavage below that of (0001) AlN. We use the pymatgen~\cite{ong2013} lattice matching algorithm to epitaxially match the perovskite to the layers extracted in our search. We identify 42 cleavage surfaces as potential substrates when using the screening criteria of (i) a work of cleavage less than that of (0001) AlN, (ii) an induced strain energy below 2~meV/atom, (iii) no {\it f}-valence species, and (iv) epitaxial matches to a single unit cell of BaSnO$_3$. Substrate matching also requires considering the chemical compatibility between the substrate and the thin film, and thus we focus further discussion on substrates with chemically inert surfaces.

We highlight here three potential substrates: (001) Rb$_2$NiO$_2$,~\cite{rieck1973oxoniccolate} (001) NiO,~\cite{rezende2001} and (001) CaSe~\cite{kohle1977}. The work of cleavage for each substrate is small, being 19, 72, and 56~meV/\AA$^2$, respectively. The resulting epitaxial strain of BaSnO$_3$ for Rb$_2$NiO$_2$ and NiO are also small, with only +0.3\% and +0.4\%, which correspond to strain energies of 0.2 and 0.4~meV/atom, respectively. These are an order of magnitude smaller than currently used substrates such as (001) SrTiO$_3$ ($-5.4$\%) and (001) MgO (+2.2\%) indicating the opportunity for defect-free growth of BaSnO$_3$ on these new substrates. The strain energy when using CaSe is larger at 1.3~meV/atom, though still with a low lattice mismatch of +0.7\%.
All three precursor materials -- Rb$_2$NiO$_2$, NiO, and CaSe -- have been experimentally synthesized~\cite{rieck1973oxoniccolate,rezende2001,kohle1977}, providing new opportunities for the growth of BaSnO$_3$. In addition, the former two being oxides indicates that the surfaces will be fairly inert and unlikely to form strong chemical bonds with BaSnO$_3$.

To facilitate the use of these substrates for further studies and future synthesis efforts of epitaxial single-crystal thin films, we provide the substrate structures and the data on their stability under an open-source license at MaterialsWeb.org. Furthermore, we make the software for cleaving crystal structures available as part of the open-source MPInterfaces package~\cite{MPInterfaces, MPInterfaces-Github}.

\section{Summary}
In conclusion, we developed a data mining approach that systematically breaks bonds in three-dimensional crystals to identify cleavage planes for substrate synthesis. We identify 4,693 unique cleavage surfaces across 2,133 periodic crystals and determine their structure and stability. We show that 3,991 surfaces display a work of cleavage comparable to that of the known substrate material (0001) AlN. These cleavage surfaces show a broad distribution of electronic properties and lattice parameters. We illustrate their utility by identifying 42 substrates for BaSnO$_3$ with epitaxial matches that are an order of magnitude better than currently used substrates, explicitly discussing two oxide and one chalcogenide substrate. Though we limited our search to small primitive cells, the low surface energy of Ga$_{2}$O$_{3}$ indicates that there are many more low-energy substrates which can be cleaved from periodic crystals.

\section{Methods}

{\it Characterization of cleaved materials.} To calculate the stability of the cleaved surfaces, we perform density functional theory (DFT) calculations with the projector-augmented wave (PAW) method as implemented in the VASP package~\cite{kresse1996vasp, bloche1994paw}. The choice of PAW potentials follows the recommendation of pymatgen~\cite{jain2013}. We employ the Perdew-Burke-Ernzerhof (PBE)~\cite{perdew1996pbe} approximation for the exchange-correlation functionals. To obtain convergence of the energy to 1~meV/atom, we use a $\Gamma$-centered {\it k}-point mesh with a density of 60~$k$-points per \AA$^{-1}$ and a cutoff energy for the plane-wave basis set of 600~eV. For the cleaved slabs, we employ a vacuum spacing of at least 14~\AA\ and reduce the number of {\it k}-points in the out-of-plane direction to 1. The Brillouin zone integration uses Gaussian smearing with a width of 0.03~eV. The DFT calculations are performed spin-polarized with an initial ferromagnetic configuration. Transition metal and $f$-valence atoms are initialized with a magnetic moment of 6~$\mu_\mathrm{bohr}$ and all others with a moment of 0.5~$\mu_\mathrm{Bohr}$. We perform single-point calculations on the bulk precursors sourced from the Materials Project database to ensure accurate energy and consistency of calculation settings across all systems. We allow a maximum of ten iterations (ionic steps) during the optimization of the atomic positions in the slabs, and as not all surfaces converged within that window, we refer to this energy as ``partially-relaxed.'' For the systems that reached ten iterations, the average and median surface energy gained are 1.6 and 0.2~meV/\AA$^{2}$, respectively, from the relaxations. We keep lattice constants of the slabs fixed at the cleaved values for these calculations.

{\it Workflow.} We use the software packages pymatgen~\cite{ong2013} and MPInterfaces~\cite{MPInterfaces} to prepare the input files, organize the results for the cleavage algorithm, and analyze the results of the DFT calculations. We use the pymatgen structure matching algorithm with an atomic position tolerance of $10^{-4}$ and not permitting any primitive cell reduction, lattice scaling, or supercell transformations~\cite{jain2013}. Decreasing the tolerance to $10^{-5}$ only marginally increases the number of unique surfaces by 27, indicating that the choice of $10^{-4}$ provides high confidence in the uniqueness of the identified substrates. Any surfaces which diverged in energy or displayed a final surface energy greater than half the work of cleavage are excluded from the analysis.

\section{Data Availability}

The database of substrates identified in this work is freely available at MaterialsWeb.org. The software we developed and used to search for novel substrates is freely available as part of the MPInterfaces software package~\cite{MPInterfaces-Github}.

\begin{acknowledgments}
We thank S. Xie, A. M. Z. Tan, W. DeBenedetti, I. Bazarov, and S. Karkare for helpful discussions. This work was supported by the National Science Foundation under
grants Nos. PHY-1549132, the Center for Bright Beams, DMR-1542776 and OAC-1740251, and the UF Informatics Institute. This research used computational resources of the University of Florida Research Computing Center. Part of this research was performed while the author was visiting the Institute for Pure and Applied Mathematics (IPAM), which is supported by the National Science Foundation under grant No. DMS-1440415.
\end{acknowledgments}
\vspace*{1.8em}

\section*{Competing Interests}

The Authors declare no Competing Financial or Non-Financial Interests. \vspace*{0.2em}

\section*{Author Contribution}

All authors contributed extensively to the work presented in this paper. J.T.P. and R.G.H. conceived the cleavage algorithm and structure search strategy. J.T.P. implemented the algorithm and performed the structure search and analysis. A.G. devised the application of the substrate database to materials and compiled the list of commercially available substrates. A.G., C.P., and K.S contributed the list of commercially available substrates. J.T.P. created the structure files for these commercial substrates. J.T.P., A.G., and R.G.H. contributed to the writing of the manuscript.

%

\end{document}